\documentclass[a4paper,11pt]{article}

\usepackage{jinstpub} 

\usepackage{lineno}

\title{\boldmath Drive beam sources and longitudinal shaping techniques for beam driven accelerators}

\author[a]{F. Lemery,}
\author[b]{G. Andonian,}
\author[c]{S. Doebert,}
\author[d]{G. Ha,}
\author[d, e]{X. Lu,}
\author[d]{J. Power,}
\author[d]{E. Wisniewski}

\affiliation[a]{Deutsches Elektronen-Synchrotron DESY, Notkestr. 85, 22607 Hamburg, Germany}
\affiliation[b]{University of California Los Angeles, Los Angeles, CA 90095, USA}
\affiliation[c]{BE Department, CERN, Geneva, CH-1211, Switzerland}
\affiliation[d]{Argonne National Laboratory, Lemont, IL 60439, USA}
\affiliation[e]{Northern Illinois University, DeKalb, IL 60115, USA}

\emailAdd{francois.lemery@desy.de, gerard@physics.ucla.edu, steffen.doebert@cern.ch, gwanghui.ha@gmail.com, xylu@niu.edu, JP@anl.gov, ewisniew@anl.gov}

\abstract{Linear colliders are an attractive platform to explore high-precision physics of newly discovered particles. The recent significant progress in advanced accelerator technologies has motivated their applications to colliders which has been discussed in the {\sc alegro} workshop. In this paper we discuss structure wakefield acceleration, namely collinear wakefield acceleration and two-beam acceleration. We especially discuss available drive and witness beam sources based on L and S-band radiofrequency technology, and also summarize available and forthcoming longitudinal shaping techniques to improve the overall acceleration efficiency via the transformer ratio.}


\proceeding{ICFA Beam Dynamics Newsletter, Issue 83}

\begin{document}
\maketitle
\flushbottom

\section{Introduction}
\label{sec:intro}

Colliders have revolutionized fundamental physics by unveiling the resonant building blocks and forces of nature. Colliders take two general designs, in circular colliders, the particles are recirculated which significantly reduces the radiofrequency power needed to accelerate the beam compared to linear colliders. However, at higher energies, synchrotron radiation losses from leptons limit the beam energies to $\mathcal{O}$(100~GeV). This limitation has motivated the use of more massive hadrons whose beam energies are limited by the ring circumference and magnet strengths, see e.g. the large hadron collider (LHC) with a nominal design beam energy of 7~TeV~\cite{LHC}.

In contrast, linear colliders avoid synchrotron losses and therefore beam energies are limited by the acceleration gradient and length of the linacs.  This is especially appealing for lepton  colliders which provide significantly cleaner signals compared to hadrons.  However, with the Higgs mass of $\sim$125~GeV, and interest in $t\Bar{t}$ production, a minimum center of mass energy of $\sim$360~GeV is favorable. Moreover there is significant theoretical interest in exploring much larger energy scales beyond 3 TeV. The development of a linear collider based on conventional radiofrequency technology would therefore have a substantial footprint limited by the accelerating gradients of approximately 20~MV/m.  

Two linear colliders have been proposed on a high level, the international linear collider (ILC)~\cite{ILCTDR} aims to deliver an efficient superconducting approach, while the compact linear collider (CLIC)~\cite{CLICCDR} proposes a large accelerating gradient (100~MV/m) using a two-beam, high current approach.  Alternative colliders with large accelerating gradients have been discussed e.g. plasma, dielectric-laser, or wakefield approaches in the {\sc alegro} workshop~\cite{alegro}.  In this paper we discuss collinear wakefield acceleration (CWA) for collider applications~\cite{vossWeiland}. 

In CWA, a drive beam creates a wakefield in a high-impedance medium, e.g. dielectric-lined~\cite{dielectric}, corrugated~\cite{corrugated}, or bimetalic waveguides~\cite{bimetalic}, and plasmas (although our discussion here does not include plasmas, the principle is very similar).  A trailing and properly delayed ``witness'' or ``main'' beam is accelerated in the wake. Contrary to two beam acceleration, here the drive and witness beams use the same beamline which facilitates energy transfer between the beams. However, this also increases challenges in beam transport due to the relatively large differences e.g. charge and energy, between the two beams. Wakefields generated in Cherenkov waveguides have also been discussed to support THz pump sources at XFELs, for a comprehensive discussion on the wakefields produced in these Cherenkov waveguides, see \cite{floettmann}.  In addition, we note that recently a technique demonstrated a way to characterize the dispersion relation of these waveguides by employing a moving obstacle in conjunction with a vector network analyzer, see~\cite{kellermeier} for details.

Longitudinally shaping the drive beam in CWA can significantly improve the efficiency of the scheme.  In an ideal configuration the longitudinal current profile can produce a nearly uniform decelerating beam over the bunch. This technique allows for a relatively small decelerating field over the drive beam while supporting a larger accelerating field for the witness bunch.  The ratio between the accelerating field over the decelerating field is known as the transformer ratio. Various techniques have been proposed and demonstrated for shaping the current profile, including emittance exchange~\cite{PhysRevLett.120.114801}, correlation based methods~\cite{England:2005prstab,Muggli:2010prstab,Andonian:2011apl,Piot:2012dualfreq, Shchegolkov:2015}, and laser-based techniques~\cite{lemerySmooth,pitzR}.

Linear colliders based on CWA have been proposed decades ago~\cite{CWAcollider}. However in these discussions, the transformer ratio was generally obtained through geometric methods or relied on non-conventional beam profiles, especially transverse rings ~\cite{vossWeiland,CWAGUN,Simpson}.  These conditions required more complex experimental considerations which eventually led to favor simpler conventional methods.

In this paper we review readily available drive beam sources which have matured in recent decades based on L and S-band technology.  Both technologies are appealing to support beam driven acceleration in different regimes.  L-band technology is attractive for its robustness in producing very large bunch charges $\mathcal{O}(100 nC)$ which can drive large gradients in lower-frequency structures.  While S-band in contrast, is attractive due to its significantly larger accelerating gradients $\mathcal{O}(100 MV/m)$ (on the photocathode) which can support the production of lower emittance beams which can be utilized in larger frequency wakefield structures which can support larger acceleration gradients.  We also discuss available and forthcoming longitudinal shaping techniques to enable high transformer ratio acceleration.  These techniques can be employed in both L and S-band technologies.  Finally we discuss key differences between CWA and TBA and point out challenges and issues which need to be resolved before a full concept can be realized.

\section{Drive beam sources}

\subsection{S-band}
\subsubsection{Introduction}
Drive beams for advanced wakefield acceleration concepts usually require either a high single bunch charge with a high repetition rate or a medium bunch charge within a bunch train in order to provide enough energy to accelerate the required main beam. In the case of a parallel, two-beam scheme the emittance of the drive beam might be less important than in the case of a collinear scheme where both the drive and main beams travel through the same accelerating structure ~\cite{CLICCDR}. There are mainly two types of cathodes available to serve as an electron source, the classical thermionic cathode and a photocathode. These can be combined with initial DC acceleration or with acceleration in an RF cavity ~\cite{hessler2, pepitone, chevallay2}. The thermionic injector typically starts with long bunches, low energy and needs a dedicated bunching system leading to significant emittance growth. On the other hand, such a source can deliver very high bunch charge (several nC) very reliably, with high stability and almost maintenance free ~\cite{pepitone}. If a small emittance is required for the drive beam the only choice is a RF-photo gun where a photocathode is placed in a short standing wave cavity allowing immediate acceleration and emittance preservation. Those guns can deliver nC type bunches with an emittance of the order of a few um ~\cite{mete, chevallay2}. The downside is the sensitivity of the high quantum efficiency (Qe) cathodes, which require a very good vacuum of the order of $10^{-10} mbar$ and have a limited lifetime. The charge stability of such a system is typically less stable compared to a thermionic gun ~\cite{divall}.
In the following, we will focus on RF-guns as a potential source for a collinear wakefield accelerator.

\subsubsection{S-band RF-Guns}
The majority of existing RF-guns using S-band frequencies, 2.85 GHz (American) or 3 GHz (European). A typical gun consists out of a 1.5 or 2.5 cell standing wave cavity, with a dedicated solenoidal field around it or close by to compensate the defocusing occurring during acceleration. The cathode is placed in the centre of the half-cell to profit from a high surface field for initial acceleration of the electrons liberated by the photoelectric effect. The laser employed in a photogun largely determines the initial beam parameters; e.g. the transverse beam size, longitudinal beam size,  charge, and its detailed distribution ~\cite{metethesis}. The time structure of the beam is also mostly determined by the laser. RF  cavities are conventionally made of copper and therefore the cavity wall itself can be used as a cathode. In order to use different cathode materials, in particular semi-conductors, a so called load lock system is used which allows to insert a coated copper plug into the back plane of the gun. In many cases, it is necessary to do this operation under vacuum with sophisticated vacuum transfer systems ~\cite{ganter}.

S-band guns operate with an RF pulse length of up to a few $\mu$s and a peak power of 10-15 MW in single bunch mode or in multi bunch mode. For a multi bunch operation a beam-loading compensation scheme can be used, injecting the bunch train during the ramp of the rf field. These cavities are operated mostly with a repetition rate of 50-100 Hz, although designs have been made for up to 400 Hz ~\cite{mckenzie}. Beyond these repetitions rates, these cavities are limited by average powers which can be overcome by using higher frequencies.

A typical S-band gun reaches an accelerating gradient of up to 120 MV/m allowing the acceleration of electron bunches with very small emittance growth. For FEL applications with bunch charges of the order of 100 pC, emittances in the sub-micron range have been achieved ~\cite{alesini}. A simple scaling suggest that the emittance increases with the square root of the bunch charge. Many RF designs have been made and tested around the world focusing on high gradient, high power or coupling to avoid multi-polar fields to mitigate emittance growth ~\cite{alesini, raquin, faillace, schaer, alesinicband}. Details can be found in the literature.

\subsubsection{Parameter range}
The following table \ref{paratab} shows the range of single bunch parameters, which are typically available for S-band RF-guns. Of course pushing certain parameters is at the expense of others.

\begin{table}[htbp]
\centering
\caption{\label{prartab} Range of single bunch parameters in S-band RF-guns.}
\label{paratab}
\smallskip
\begin{tabular}{lcl}
\hline
Parameter&	Typical Range&	Remarks\\
\hline
Gradient (MV/m)&	80-120& \\	
Energy (MeV)&	4-6& \\	
Charge (nC)&	0.1-10& \\	
Emittance ($\mu$m)&	0.2-20&	0.2 $\mu$m for 100pC\\
Energy spread (\%)&	0.1-1&	Depends on phase\\
Repetition Rate (Hz)&	1-100&	Up to 400 Hz in S-band\\
Bunch length (ps)&	0.2-10& \\	
\hline
\end{tabular}
\end{table}

Multi bunch operation is possible but needs special precaution for beam loading compensation and bunch-to-bunch stability ~\cite{mete}.

\subsubsection{Cathodes}
We distinguish between two main types of cathodes relevant for those applications. Metal cathodes where Copper is most common, but Magnesium and Iridium have been used as well. They reach typical quantum efficiencies in the range of $10^{-5}-10^{-4}$ but are quite robust and do not need an extremely small pressures. Obviously, in particular in the case of copper they are very straight forward to use in an RF cavity and can simply be part of the cavity wall. A limitation of metallic cathodes is their ablation threshold if high charge and small spot sizes are required.

The second type of cathodes are semi-conductors like $Cs_{2}Te$, $CsSb$, $Cs_{3}Nb$ or $GaAS$. The semiconductors are produced as a thin layer on a metallic substrate and then introduced as a part of the cavity-wall into the RF-gun. They promise much higher quantum efficiencies on the order of up to 0.1-0.2. The thin film layers are very sensitive to residual gas pressure in the gun and to particle back bombardment. Typically a static vacuum of $< 10^{-10} mbar$ is needed for those cathodes in the gun. The high quantum efficiencies have therefore a limited lifetime, which depends on the charge extracted and on the dynamic vacuum during operation. In the case of CLIC, studies have shown a two week lifetime above 0.3 Qe for a charge extraction of a nC ~\cite{chevallay, hessler}. These cathodes require a sophisticated infrastructure for fabrication as well as to transfer them under vacuum into the RF-guns for operation.

\subsubsection{Limitations, future directions}
RF guns have been studied and built as well at C-band ~\cite{alesinicband} and X-band frequencies ~\cite{limborg}. In general a higher frequency promises a higher gradient (150-200 MV/m) and therefore a better emittance as well as lower stored energy which could enable higher repetition rates. Of course, this is at the expense of a smaller gun, which complicates cooling, vacuum pumping and laser coupling. The smaller gun apertures lead as well to more wakefield effects. Nevertheless future developments in particular for light sources aim for higher frequencies to profit from a better emittance and higher repetition rates. For high bunch charge applications, those frequencies might be not so important for time being.

\subsection{L-band}
\subsubsection{Introduction}
S-band, C-band, X-band and other RF frequency options can be used in the production of high charge electron bunches in the range of 1-10 nC.  However, in the production of extremely high charge (10-100~nC) with short bunches of a few ps in length, the L-band 1.5~cell gun has advantages. First, as the charge is increased above 10~nC per bunch, space-charge forces near the cathode are intensified when the geometry restricts operations to small beam sizes at the cathode. Second, the large stored energy in the L-band gun, approximately 30~J, results in reduced beam loading compared to higher frequency guns.  We note there has been work reported on high-charge operation in S-band structures, see ~\cite{Bossart}.  However while charges of 35~nC per bunch have been generated, the beam qualities suffer significantly.  A suitable operational limit for quality bunches from S-band technology is approximately 10~nC.

\subsubsection{Why L-band for extremely high charge?}
 The very first normal conducting radio frequency (NCRF) photoinjector was L-band. That first photoinjector operated at 1.3 GHz, was developed and demonstrated at Los Alamos National Laboratory (LANL), as reported in 1988 \cite{RSheffield1}. The primary goal was to develop an electron source to produce high brightness beams for FELs. Soon after, Los Alamos produced and operated successful L-band photoinjectors such as APEX and AFEL. NCRF L-band photoinjectors have continued to develop as high-brightness electron sources~\cite{DOWELL200661}. The photoinjector development program at TTF and PITZ (DESY) has been extremely successful~\cite{Dwersteg97}, achieving high performance benchmarks when deployed at facilities such as FLASH and European-XFEL~\cite{Brinker:IPAC2016-TUOCA03}. At European-XFEL very high average current is achieved in their burst mode operation. In this mode, the European-XFEL can operate with production of up to 2700 bunches of 1 nC at 10 Hz giving a total of 27000 bunches per second. The European-XFEL design parameters are quite challenging, including a very long RF pulse approaching 1 millisecond, high average RF power dissipation, high gradient (60 MV/m) on the cathode. Developments such as implementation of coaxial rf power feed, improvements in laser technology, RF cavity cooling, photocathode technology have continued to push the desired parameters \cite{Paramonov2017}. Many other L-band guns have been or still are operating for various purposes, most with an RF pulse length of 100s of microseconds.

 In the development of extremely high charge sources, there are several important considerations. The cathode, the gun irises and gun output port all have to be over-sized to mitigate strong space-charge forces of high charge bunches since these features allow for reduced beam charge density. In other words, high charge implies big beams. Space-charge effects also dictate that the field on the cathode must be high enough to overcome space-charge limited emission and the beam energy gain must be large enough to produce highly relativistic beams from the gun very quickly (space charge forces scale directly with the charge density and inversely as the square of the relativistic Lorentz factor \cite{Wiedemann2007}).  As discussed in the previous section, higher frequency guns can support higher accelerating fields, but their short lengths result in lower energies at the gun exit. The 1.3 GHz L-band gun is large enough to accommodate large beams while also attaining high accelerating fields and a large gun exit energy\cite{Gai1998,Conde2001}.

 L-band photoinjectors for high charge, with their large cathodes and irises, allow production of bunches and bunch trains with large transverse dimensions, which are typically generated from a single laser pulse split n times producing $2^n$ bunches spaced with the L-band period. For a typical 8-bunch train generated from a 1.3 GHz gun, this results in a pulse train length of several nanoseconds. This beam current structure is desirable for generating intense, short-pulse (a few nanoseconds) RF power, producing very high gradients with reduced probability of breakdown. The large stored energy of L-band guns ($\approx$30 J), is especially important for high charge bunch trains. Another issue (discussed in later sections) is the potential for shaping (transverse and longitudinal) of the current density distributions of the bunches and the train.
 
 Finally we note that there has been work on high-charge S-band

  \subsubsection{Collinear Wakefield Accelerator schemes using extremely high charge bunch trains}
 Some designs for Beam-driven wakefield accelerator have been proposed and researched that rely on extremely high charge drive bunches [O(10–100 nC)] passing through slow-wave structures to excite high-gradient time-varying electromagnetic wakefields, thus extracting very high peak power directly from the beam.\cite{Jing2012}

 The generated RF power may be used directly within the structure to accelerate a trailing “main” bunch, a scheme called Collinear Wakefield Accelerator (CWA). Another option is to utilize the power elsewhere by coupling into a waveguide (100s of MW have been extracted, but potentially GWs), transported and then coupled out to an accelerating structure in a parallel beamline timed to accelerate the main bunch, also called a Two Beam Accelerator (TBA). CWA offers a seemingly simpler configuration, where both the drive and main bunches are transported along the same beamline, however this can create difficulties due to the fact that the two beams traversing the same lattice have very different charge/current profiles and therefore beam dynamics differences. One way this issue can be addressed is by separate drive and witness sources injected into a common structure.

\begin{figure}[htb]
   \centering
   \includegraphics*[width=.9\columnwidth]{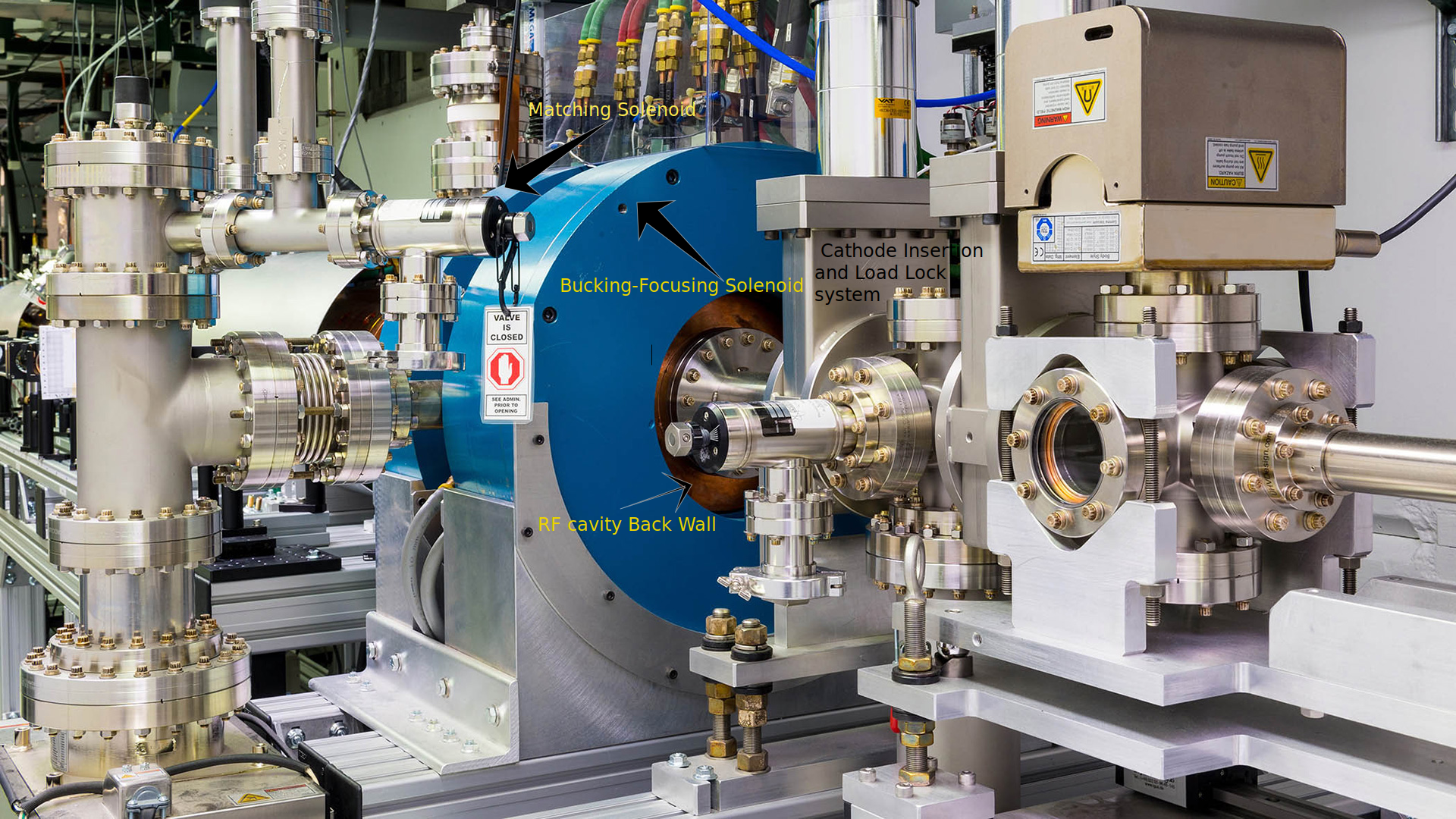}
   \caption{The current AWA drive gun: G3. Beam direction is right to left. The load-lock system attached to the back wall of the gun is used to insert the Cs2Te cathode. Solenoids (blue) for emittance compensation and RF matching.\label {fig:G3}}
\end{figure}

 \subsubsection{Argonne Wakefield Accelerator High Charge Production}
Producing extremely high charge electron bunches currently relies on L-band photoinjectors and due to the intensity of these bunches, puts stringent demands on the associated photocathode and laser system. The Argonne Wakefield Accelerator Group (AWA)  (Argonne National Laboratory (ANL), USA) currently operates two L-band photoinjectors capable of high charge generation.  These capabilities have been developed over the past 20 years as a critical part of the group's Structure Wakefield Acceleration (SWFA) research program.

There are two similarly designed L-band photoinjectors in use at AWA which are used for high charge generation for wakefield experiments, first proposed in 1997 \cite {Gai1998}. The original, designated G2 (ca. 2003), was designed to use a ``brute force'' approach to high charge production as described in \cite{Gai1998}, seeing this as a practical way forward to move beyond the 10 nC limit. The basic idea was to use a 20 mm laser spot, (cathode diameter= 31 mm), large irises: inter-cell iris diameter 5.5 cm, gun exit iris 5.0 cm, and high field on the cathode (design called for 85 MV/m but was limited in practice to about 75 MV/m), producing large high charge 7 MeV bunches out of the gun. Two opposing solenoid coils (bucking and focusing) configured with the interface of the steel at the cathode plane provide emittance compensating magnetic field with zero field on the cathode, and a large bore matching solenoid is located at the gun exit.

The first gun proved capable of 100 nC single bunches or trains of 4 bunches of 25 nC each spaced 0.77 ns apart. The primary limitation to moving to higher charge was the QE of the photocathode. The magnesium cathode was chosen for its fairly high QE plus its robustness outside vacuum, which allowed avoiding the expense and complication of a load-lock system.
In the pursuit of high charge trains, G2 was superseded by the current drive-gun G3 (pictured in  fig.~\ref{fig:G3}). The design based on the G2 design but enhanced by an AWA-developed large format (31 mm)  Cs2Te photocathode, a semiconductor thin-film  requiring a load-lock system and lots of attention to  vacuum. The AWA-produced photocathode typically has an operating QE of 2\% or higher. Because of this high QE,  G3 is capable of single bunch charge >100 nC as well as bunch trains > 500 nC. Increasing charge from the gun is not an issue. Using a plate-beamsplitter based multi-splitter, G3 has generated bunch trains consisting of of $2^n$ bunches, up to 32 bunches. The addition of a microlens array (MLA) based homogenizer has greatly improved the uniformity of the laser profile.\cite{Halavanau2016} G3 was used to do a simple staging TBA experiment producing 2 8-bunch trains to drive two separate wakefield structures powering and accelerate a witness beam produced by G2 \cite{JING201872}.  With an existing slow-wave structure, AWA routinely transmits 100\% of 400 nC in an 8-bunch train through a 17 mm aperture, 30 cm long structure generating short pulse,  400 MW RF peak power.

\subsubsection {L-band Photoinjector parameters}
Selected AWA drive gun,laser and beam parameters are summarized in Table~\ref{tab:drivegun}~,   Table~\ref{tab:drivelaser}, Table~\ref{tab:drivebeam}~\cite{Gai1997,Wisniewski2014,Wisniewski2015}.

\begin{table}[h]
 \caption{AWA L-band gun operating parameters}
\centering                          
 \begin{tabular}{ll}          
\hline\hline    
Cathode peak RF field  & 75 MV/m \\ 
\hline
RF pulse length        & 5 $\mu$s \\ 
\hline
Average dark current    & $<$5 nC/RF pulse\\
\hline
QE\%                   & 2\%\\
\hline
QE lifetime            &$>$2 years\\

\hline
Single-bunch charge     &100pC to >100nC\\
\hline
Bunch train mode parameters:       &$\bullet~$up to 70 nC/bunch \\
        &$\bullet~$2 to 32 bunches\\
        &$\bullet~$bunch spacing = 769 ps\\
        &$\bullet~$Max charge = 0.6 $\mu$C\\
\hline                                   
 \end{tabular}
 \label{tab:drivegun}
 \end{table}
 
 \begin{table}[h]
 \caption{AWA L-band gun laser parameters}
\centering                       
 \begin{tabular}{ll}       
\hline\hline    
Laser wavelength  & 262 nm \\ 
\hline
Laser pulse length      & 0.3-6 ps FWHM \\ 
\hline
laser energy per nC   & 200 nJ/nC \\

\hline
 maximum laser spot size & 22 mm \\
\hline

\hline                              
 \end{tabular}
 \label{tab:drivelaser}
 \end{table}
 
 \begin{table}[h]
 \caption{AWA G3 photoinjector selected high charge beam parameters (from simulations) see Fig. \ref{fig:BeamsimsAWA}}
\centering                          
 \begin{tabular}{ll}          
\hline\hline    
Charge per bunch & 50 nC \\
\hline
 laser spot size & 18-22  mm \\

\hline
bunch length      & 0.8 - 2.5 mm \\ 

\hline
normalized transverse  emittance  & 100 micron \\

\hline

\hline                                   
 \end{tabular}
 \label{tab:drivebeam}
\end{table}

\begin{figure}[htb]
   \centering
   \includegraphics*[width=.8\columnwidth]{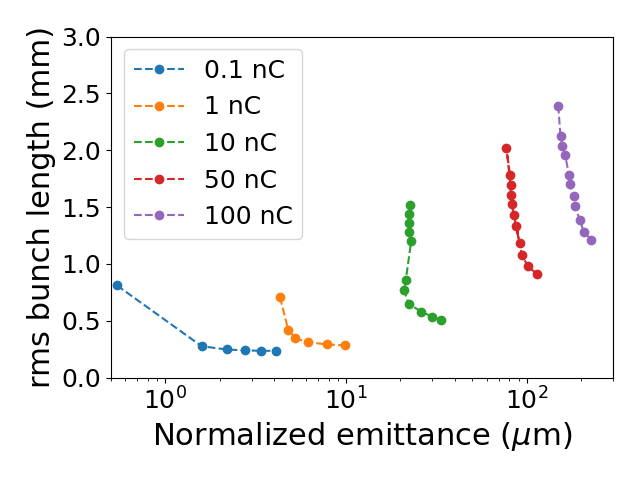}
   \caption{AWA L-band gun simulation results for a range of charges (courtesy G.Ha).\label{fig:BeamsimsAWA}}
\end{figure}

 The beam simulations results plot (Fig. \ref{fig:BeamsimsAWA}) cover a fairly wide dynamic range of single-bunch charge, 0.1 nC to 100 pC which serves a wide-range of experiments for the research program. With the high-medium-low charge and shaping capabilities, it might be said that the AWA experimental program also has a wide dynamic range. Results plotted are from GPT, courtesy of G.Ha.

\subsubsection {CWA experiments at AWA}

AWA has a long history of CWA research, here we present a few recent highlights.The AWA drive beam source feeds into a switchyard providing beam to either the straight-ahead  experimental area or else to the Emittance Exchange (EEX) beamline which provides beam-shaping capabilities based on phase-space manipulation (discussed in a later section). AWA has performed significant CWA experiments recently including  the Observation of High Transformer Ratio in a Dielectric Slab structure using trailing witness beam\cite{PhysRevLett.120.114801}, CWA PWFA using a shaped bunch for high transformer ratio and trailing witness beam (in collaboration with UCLA) \cite{PhysRevLett.124.044802}, as well as a CWA THz metallic structure driven by a high-frequency bunch-train in collaboration with Pohang accelerator laboratory (PAL)  (to be published).

\subsubsection{L-band high charge source - improvements}

It seems that the L-band high charge photoinjector has been under-researched, so there is probably a large amount of room to improve. Indeed, there are several areas that come to mind.  These include, but are not limited to, the following:

\begin {itemize}
\item the average current for high-charge production can be improved. To begin with, G3 rep rate is low (2 Hz) and the duty factor is 1, increasing one or both of these would definitely be required for a practical machine.
\item RF power requirements
\item higher gradient
\item charge balance along the train (currently less than 80\%)
\item Longitudinal laser shaping
\end {itemize}

The G3 rep rate is rather low (2 Hz) and the duty factor is only 1, increasing one or both of these would definitely be required for a practical machine.  In the near future AWA plans to test an increase in duty factor to 2 trains delivered per RF pulse, doubling the delivered beam current. The current machine has the capability to operate with rep-rate up to 10 Hz. This however has not been tested and will probably have to wait for some infrastructure improvements at AWA (two aging L-band klystrons need to be replaced, for example). The infrastructure improvements could also assist with the high RF power requirement to achieve higher gradients. Up to 100 MV/m gradient on the cathode is possible and has been demonstrated but the current gun design has a prohibitively high klystron power requirement for practical operation at this level.

Improvements in charge balance along the train (currently less than 80\% due to limitations of the optical components) will directly benefit the SWFA experiments relying on high-charge drive trains. This issue is currently being researched and addressed at AWA. Development of an improved multi-splitter design is in progress.

Longitudinal laser shaping could bring improvements in the beam emittance and help reduce the effects of space-charge driven emittance growth, improving the charge production and transport capability. Some methods have been suggested and tested for the 1- 10 nC level,~\cite{Beaudoin2016}. These methods (and others) should be modeled, then if possible,  implemented and tested  for extremely high charge. 

Tackling these issues (and others) can lead to higher average current, brighter beams, higher transformer ratio, as well as improved beam transmission and stability.  It appears that continuing the development of a range of RF frequency (S-band, C-band, L-band and others) drive-beam sources will be quite beneficial, as they have been demonstrated to serve different requirements in different ways. L-band sources provide some unique qualities (in particular, higher charge per short bunch) to help further CWA research toward an efficient future collider concept.

\section{Longitudinal shaping}
Longitudinal shaping is complimentary technique which can significantly improve the overall efficiency of CWA.  In this section we discuss the various available and emerging longitudinal shaping techniques which can be used to provide high transformer ratio acceleration.

\subsection{EEX-based shaping}
Emittance exchange (EEX) is a method to exchange one of the transverse phase spaces and longitudinal phase space \cite{cornnacchia-2002-a,emma-2006-a}. EEX was proposed to achieve a low transverse emittance with a high peak current for light sources \cite{cornnacchia-2002-a}. Later it was also considered a method to obviate the damping ring from linear colliders \cite{emma-2006-a}. While these original applications focused on the exchange of emittances, its other applications that exploit the exchange of phase spaces studied actively in the last decade \cite{ha-2019-a}. Longitudinal profile shaping was one of those applications.

\begin{figure}
\centerline{\includegraphics[width=1.0\textwidth, keepaspectratio=true]{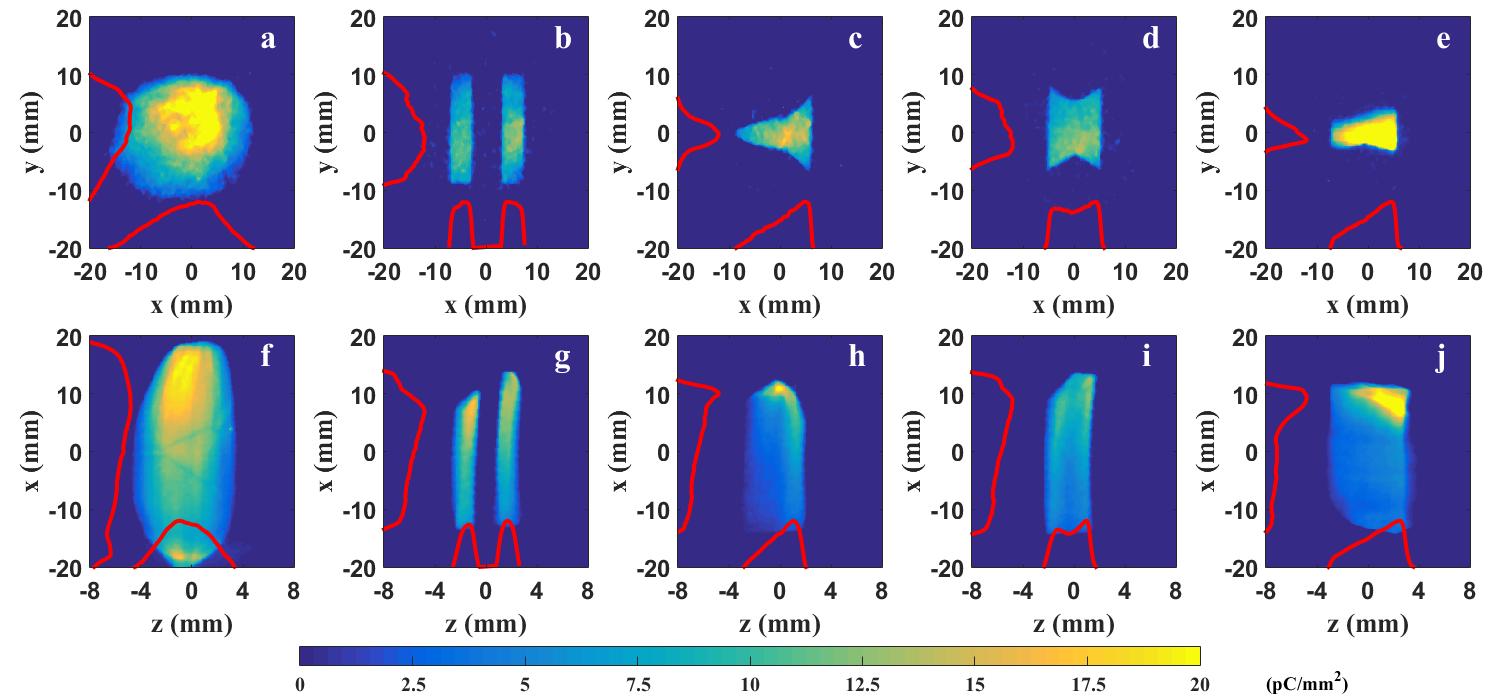}}
\caption{Longitudinal shaping by EEX beamline. The top row shows beam's $x$-$y$ image after the mask but before the exchange. The bottom row shows corresponding beam's $x$-$z$ image after the exchange. Figure is from Ref.~\cite{ha-2017-a}}\label{fig_sec32_shaping}
\end{figure}

Longitudinal shaping using an EEX beamline was firstly proposed in 2011 \cite{piot-2011-a}. Due to the exchange of phase spaces, the particle’s final longitudinal coordinates are governed by their initial transverse coordinates.
\begin{equation}\label{eq_sec32_eex}
    z_f = R_{11} x_i + R_{12} x'_i
\end{equation}
where $R_{ij}$ is $(i,j)$ element of a beamline's transfer matrix.
In other words, EEX beamline enables the direct control of longitudinal phase space via any transverse manipulations such as collimation or nonlinear magnets. The easiest way to achieve an arbitrary profile is using a transverse mask. This method was experimentally demonstrated in 2017 \cite{ha-2017-a}. During the experiment, several masks having different shape was applied to the beam to shape its transverse profile. Then, the following EEX beamline converted the shaped profile to the longitudinal profile (see Fig.~\ref{fig_sec32_shaping}). This method was used later to demonstrate transformer ratio enhancement for both SWFA \cite{PhysRevLett.120.114801} and PWFA \cite{PhysRevLett.124.044802}. Both experiments achieved the current world record transformer ratio for each acceleration scheme. Although the experiment generated triangular profiles, the method can generate any complex profiles that Ref. \cite{bane-1985-a} and \cite{lemery-2015-a} proposed, and reverse-triangle for main bunch’s beam loading control, which are key to enhance the acceleration efficiency.

As we can see from Eq~\ref{eq_sec32_eex}, the method directly controls the longitudinal phase space. Thus, in the ideal situation (i.e., all limiting factors are properly suppressed), this is a method that can provide extremely high shaping accuracy (e.g., hard-edge tail). At the same time, the method does not require a specific location for the shaping. Thus, it can shape both the drive bunch from the injector and the main bunch after the damping ring. Also, the same principle works for positive charges, so the EEX-based shaping can shape positron bunches.

It is worth to note that there is an effort to substitute a large damping ring to a linac by using flat beam transform and emittance exchange \cite{emma-2006-a,kuriki-2016-a}. In this case, EEX is necessary component to satisfy the emittance requirement of the main bunch. This EEX beamline can provide the shaping of the main bunch simultaneously. If flat beam transform techniques for a positron beam are developed in the future, EEX would support both the damping-ring-free feature and the shaping for the positron beam.

Although the method is a strong candidate as a shaper for future accelerators, it still has three major challenges to overcome. First, the shaping was only demonstrated for a few nanocoulomb bunches so far. All three experiments \cite{ha-2017-a,PhysRevLett.120.114801,PhysRevLett.124.044802} generated 1-2 nC triangular bunches. Previous simulation studies \cite{ha-2016-a} showed significant decreases in the shaping quality and emittance growth for higher charges due to coherent synchrotron radiation (CSR) effects. However, 10 nC or beyond is necessary to achieve GV/m-class accelerating gradients. Second, the shaping using a mask introduces significant charge loss. Typical transmission for a triangular bunch is 25-50\%. Such a charge loss may introduce thermal or radiation issues which would need to be handled especially for large average powers. Also, the beam’s emittance would be another issue related to the charge loss. Third, most injectors provide a small transverse emittance and a large longitudinal emittance. Thus, a single exchange results in a large transverse emittance which is not preferred for any applications. Also, initial timing or energy jitters become transverse position and angle jitters after the exchange. Such jitters may introduce emittance growth or instabilities.

Various efforts are ongoing to resolve those issues. For example, an asymmetric double dogleg beamline was proposed to mitigate CSR effects on the shaping process \cite{ha-2016-a}. Reducing R56 of the second dogleg mitigated the CSR effect on the shaping significantly, and the simulation showed a high-quality 11-nC triangular profile. Also, several different methods (e.g., shielding) have been explored to mitigate emittance growth by CSR in the EEX beamline \cite{ha-2017-b}. Regarding charge loss, nonlinear magnet or transverse wigglers can control x-x’ or y-y’ correlations, which eventually controls the beam’s transverse profile \cite{yuri-2007-a,ha-2019-b}. These methods are attractive lossless shaping options. A double EEX beamline \cite{zholents-2011-a}, which is a series of two EEX beamlines, is considered as one of the solutions to avoid issues from the exchange itself \cite{ha-2017-c}. The shaping can be achieved in the same manner at the middle section in-between two EEX beamlines while the original transverse emittance returns to the transverse plane after two EEXs.

\subsection{Correlation based shaping methods}

A class of beam tailoring techniques referred to as correlation-based, accomplish longitudinal shaping by the introduction, and manipulation, of one-degree of freedom in the bunch envelope phase space. 
By exploiting correlations between the different coordinates, shaping is performed in a readily accessible dimension, then transformed into the longitudinal dimension.
In general, correlation-based shaping incorporates a modulation of the beam energy, followed by a conversion of the modulation into time.
The longitudinal coordinate of a particle in the bunch can be written as
the transformation from initial, $z_0$, to final, $z_f$,  as 
$z_f = z_0 + R_{56} \delta$
to first order in correlated energy spread $\delta$, where $R_{56} = \frac{\partial z}{\partial \delta}$ is known as the longitudinal dispersion, or correlation coefficient.
The beam energy modulation can be performed by many methods, including both by external sources or internally generated fields, while the longitudinal conversion can be accomplished by drifts for low energy beams, and magnetic dispersion beamlines, such as chicanes and doglegs, for high energy beams.
This section describes some of the ways to a modulate and convert the beam phase space that have recently been demonstrated in experiment.
The examples include using intercepting masks in high dispersion beamlines, using multiple frequency linacs to impart nonlinear modulations, higher order manipulation in longitudinal dispersion, and self-generated modulations from wakefields.

\subsubsection{Intercepting masks in dispersive beamlines}

Temporal shaping using masking requires the introduction of a correlation between the beam transverse and longitudinal dimensions, such that the mask features are reshaped on the beam current profile.
The main advantage of masking techniques is that transverse shaping is relatively straightforward with standard materials, and present-day machining yields high precision for sharp features in mask fabrication.

In this method, an intercepting mask with a specified shape, or transverse transmission function, is placed within the beam trajectory. 
Particles within the open aperture travel uninhibited, while the others intercept the mask
(the mask can also simply spoil the beam emittance of intercepted particles, not necessarily block transmission).
The correlation between the transverse (e.g. horizontal) and longitudinal dimensions is accomplished on a beam line with high magnetic dispersion, such as a chicane or dog-leg.
The mask is placed at a point along the dispersive beamline with high dispersion and low beta function such that the beam size is dominated by the dispersion term.
For a beam with an initial chirp (energy-time correlation) that travels through the dispersive line, at the mask location, the mask transmission function is correlated to the beam energy.
When the dispersion along the beamline returns to zero, the beam phase space is recombined and the transmitted beam transverse coordinate is now mapped onto the longitudinal profile.

\begin{figure}[h]
    \centering
    \includegraphics[scale=.75]{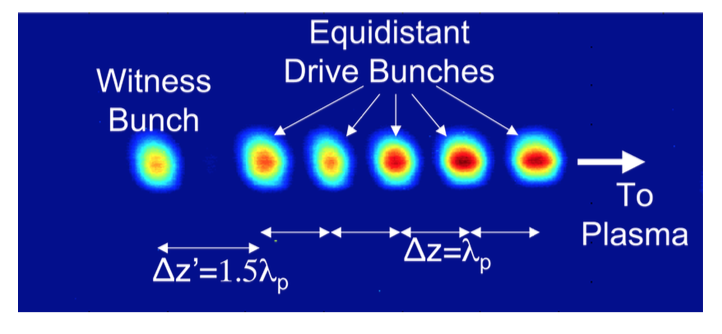}
    \caption{False-color image of bunch train used for PWFA experiments from masking method at BNL ATF. Adapted from Ref.~\cite{Muggli:2010prstab}.}
    \label{fig:muggli_bunches}
\end{figure}

An experiment to demonstrate the masking technique was reported in Ref.~\cite{Muggli:2010prstab}, which took place at the Brookhaven National Laboratory Accelerator Test Facility (BNL ATF).
The motivation was to generate a train of bunches with sub-picosecond spacing to resonantly drive plasma wakefield acceleration (PWFA).
The mask for bunch train generation is composed of an array of thin metallic wires stretched in parallel over a round aperture. 
Shaped slits cut into metallic disks can be used in place of the wires to achieve the same result.
The unspoiled portions of the beam through the mask compose the time slices of the bunch train.
The BNL ATF dogleg beamline was used to impart the necessary dispersion.
Quadrupole magnets on the dispersive line  are adjusted for high dispersion, and low horizontal beta function.
In the experiment, the electron beam had a central energy of 58~MeV and a fractional energy spread that is variable from 0.5$\%$~-~3.5$\%$, with a bunch length of 5.5~ps.
The dogleg longitudinal dispersion was $R_{56}=4$~cm.
The beam energy chirp imparted by the linac corresponds to an energy-horizontal position correlation in the dogleg, where the wire mesh is placed.
The bunch time structure is measured downstream of the dogleg using coherent transition radiation (CTR) interferometry.
The resultant measurements demonstrated bunch trains with spacing of 216$\mu$m-434$\mu$m for various energy chirp (Fig.~\ref{fig:muggli_bunches}).

The number of bunches and bunch length is controllable by varying the central energy and energy spread of the beam, or by changing the shape of the mask.
In another experiment at the BNL ATF, the masking technique was adopted to generate variable bunch trains to resonantly drive dielectric wakefield acceleration (DWA) \cite{Andonian:2011apl}.
By varying the correlated energy spread of the beam, the bunch train spacing is tunable. 
When the appropriately spaced bunch trains pass through the DWA, they selectively drive distinct modes, while others are suppressed. 
The spectral characterization of the emitted radiation shows preferential excitation of modes when the bunch spacing is equal to the resonant wavelength.
These measurements demonstrate the flexibility provided by the masking technique.

The masking method can also be used for sophisticated shaping beyond bunch trains, such as current profiles that are linearly ramped, or triangle shaped, by using a different shape mask instead of slits or wires. 
Bunches with triangle and double-triangle current profiles, provide acceleration with high transformer ratio in colinear wakefield accelerator schemes \cite{bane-1985-a,Lemery:2015prab}.
The ramped beam shaping was demonstrated in \cite{Shchegolkov:2015} at the BNL ATF where different shapes were cut in the intercepting masks to produce the desired current profiles. 
In  the experiment, different masks were used to create single triangle and double-triangle shapes, which were then sent through a DWA structure.

The advantages of the masking technique include its simplicity, as there is no need for additional power sources. 
Very high precision features are achievable, and various shapes are possible with simple mask swapping, or use of a variable mask \cite{Majernik:2021}.
In addition, most accelerator facilities already employ doglegs to branch off beamlines, or chicanes for bunch compression or delay lines, so the incorporation of bunch shaping by masking does not require major facility modification.
The main drawback, like any beam intercepting technique, is that there is an inherent loss of charge at the mask location, limiting the resultant charge available at a downstream location. 
Also since the mask must be positioned at a location of high dispersion in the beamline, coherent synchrotron radiation (CSR) effects from bends must be considered, which can lead to emittance degradation and energy spread effects.

\subsubsection{RF Modulation with longitudinal dispersion}

In the general case of longitudinal shaping, a beam chirp is imparted by an RF linac, which then undergoes conversion to the temporal coordinate in a longitudinally dispersive beamline element. Written to second-order, the transformation of the longitudinal coordinate follows as 
$z_f = z_0 + R_{56} \delta + T_{566} \delta^2$, where $T_{566}=\frac{\partial^2 z}{\partial \delta^2}$.
In an experiment conducted at the UCLA Neptune Laboratory, the second order effects on beam modulation were examined and manipulated to generate a variety of shapes \cite{England:2005prstab}.
The energy chirp imparted on the beam is proportional to the linac drive voltage 
$V(z)=V_0\cos(k z + \phi)$, where $k$ is the wavenumber associated with the fundamental frequency and $\phi$ is the phase.
The energy-time correlation in general displays some nonlinear curvature
and the second order terms can be made large to further compress the beam, as seen in Ref.~\cite{Murokh:2003vis}.
\begin{figure}[h]
    \centering
    \includegraphics[trim={0 270 0
    0},clip]{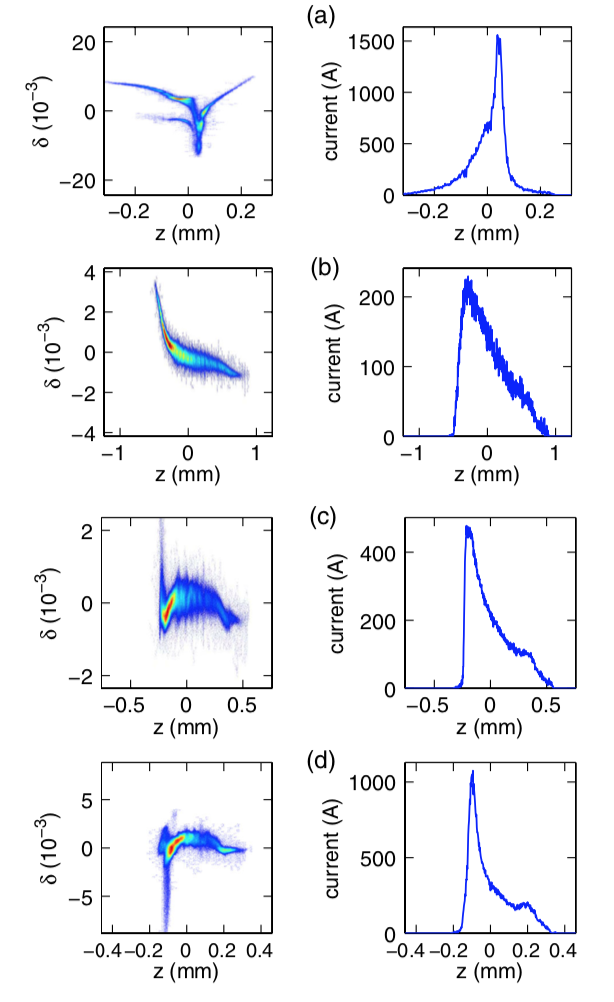}
    \caption{Samples of different longitudinal phase space (left column) and the corresponding current profiles (right column) for various settings in the dual linac and chicane shaping scheme. Adapted from \cite{Piot:2012dualfreq}.}
    \label{fig:piot_shapes}
\end{figure}

However, the higher order terms can be controlled by the use of nonlinear magnets. In the experiment, sextupole magnets were placed near high dispersion points along the beamline, to change the linearization of the dispersion term. 
The bunch shape was inferred from an analysis of the coherent transition radiation emitted as the bunch strikes a target.
A variety of $T_{566}$ values were obtained by varying the sextupole gradients, which led to a range of current profiles, including ramped beams.
Similar implementation of sextupole gradients was used to linearize the bunch train spacing for DWA experiments in BNL ATF \cite{Andonian:2011apl}

Improving longitudinal shaping features and tunability is  possible by adding additional frequency linacs to further modify the beam energy chirp.
In an experiment conducted at FLASH, a dual frequency superconducting linac was used to impose a nonlinear modulation on the beam \cite{Piot:2012dualfreq}.
When multiple linacs of different frequency are used, the energy change is proportional to $V(z)=V_1\cos(k_1 z + \phi_1) + V_n\cos(k_n z + \phi_n)$, where $k_n$ and $\phi_n$ correspond to wavenumber and phase associated with the $n$-th higher harmonic of the fundamental frequency.
In the experiment, the electron bunch was accelerated in a 1.3 and 3.9~GHz linac structure to $\sim$700~MeV final energy, then passed through a chicane bunch compressor with $R_{56}$=176mm.
The linac settings were optimized to form a linear current profile, with precision control over the shaping \ref{fig:piot_shapes}.
More exotic shapes were also produced by varying the linac phase terms for each frequency. 

The technique is readily extendable to multiple frequencies for even more exotic shaping capabilities, however the method requires the use of externally powered sources and dedicated use beamlines to achieve the desired current profile for applications.  We note this technique has also been used at MAX IV~\cite{Svensson}.

\subsubsection{Self-generated modulation from wakefields}

An extension to the energy modulation and conversion concept is the use of self-generated wakefields.
Wakefields provide the time-energy correlation on an incoming beam, without the need of powering external fields.
Wakefields can be generated in different types of media, including dielectric lined waveguides, corrugated metal structures, and plasma.
The benefits of using self-generated wakefields in structures for energy modulation include compactness and no loss of charge, unlike in masking techniques. 
In addition, since the structure can be small in size and does not require external power feedthroughs, it can be located very close to the interaction point, reducing transport effects.
The wakefield frequency depends on the material and thickness in dielectric structures, and feature dimensions for corrugated structures, providing flexibility for desired applications.
Once an energy modulation is imparted on the beam, the temporal conversion is accomplished by the dispersive term, 
$R_{56}$, which can come from a chicane or a drift length.

In two experiments at the BNL ATF, the wakefield modulation concept was tested for both bunch train generation \cite{Antipov:2013bunchtrn} and for ramped beam shaping \cite{Andonian:2017prl}.
In the former case, a strong energy modulation is imparted on an electron bunch passing through a dielectric-lined waveguide. 
The structure dimensions are chosen such that the beam bunch length is much larger than fundamental wavelength of the excited wakefield ($\sigma_z>>\lambda$). For this case the fundamental frequency was 0.805~THz.
In this regime, the bunch samples many periods of the wakefield. 
The longitudinal dispersion is provided by a chicane, that consists of 4 permanent magnet dipoles, with $R_{56}=4.9cm$.
For a fixed $R_{56}$ value, the bunch train spacing can be adjusted by changing the value of the chirp of the incoming beam.
In the experiment, the bunch train frequency spacing after the chicane was varied from 0.68 to 0.9 THz, as measured by analysis of the coherent transition radiation emitted when the bunch strikes a target.

\begin{figure}[h]
    \centering
    \includegraphics[trim={10 0 0 10} ,clip]{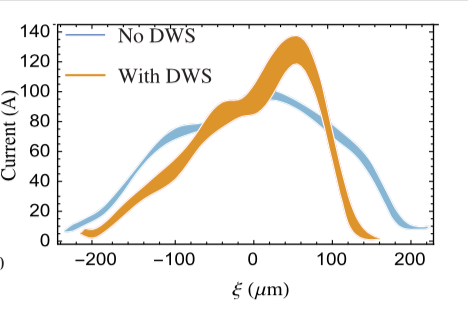}
    \caption{Ramped current profile generated from wakefield modulation compared to unperturbed beam. Adapted from \cite{Andonian:2017prl}.}
    \label{fig:andonian_ramp}
\end{figure}

In the second experiment \cite{Andonian:2017prl}, a similar technique was used to shape the bunch into a near-linear ramp, or triangle, shaped.
For this type of shaping, the dielectric structure parameters are chosen such that the fundamental wavelength is on the order of the beam bunch length ($\sigma_z\approx\lambda$).
In this regime, the bunch only partially samples a portion of the wakefield period, imparting a nonlinear energy correlation on the beam, as the energy modulation is proportional to the convolution of the wake function with the bunch profile.
Specifically, for this case, the fundamental frequency of the structure was 0.39~THz, corresponding to $\sim$1/4 of the bunch length.
After passing through the chicane of $R_{56}$=9~mm, the ramped bunch profiles were measured using both coherent transition radiation interferometry and a transverse deflecting cavity (Fig.~\ref{fig:andonian_ramp}).

For low energy beams, once an energy modulation is imparted by the wakefield, the temporal conversion can take place in a simple drift, which has an inherent $R_{56}= -L/\gamma^2$
\cite{Lemery:2014ballistic}. 
 Experimentally, this technique was demonstrated in Ref.~\cite{Lemery:2019prl}, for bunch train generation (when the bunch length is much greater than the excited wakefield period) and is extendable to triangle shaping, by choosing a medium with a suitable wakefield frequency. 

Self-generated wakefields for longitudinal shaping have advantages due to cost and size, as well as not needing an external power source. However, there are some issues that require attention when using these passive sources, including taking care to not accumulate too much energy spread during the modulation. In addition, the excitation of transverse wakefields for off-axis particles can lead to beam break up or emittance growth. Furthermore, the wakefield always acts to decrease the beam energy at the onset, which implies that shaping can only occur in one direction. However, the technique is extendable to multiple frequency structures for higher precision features, and can also be combined with the masking or EEX to improve efficiency in beam shaping.

\subsection{Laser based}

Laser based shaping approaches are also attractive, providing means to produce longitudinally shaped bunches directly from the photocathode.  A motivation regarding this concept was provided by Laziev et. al.~\cite{Laziev}, where the use of an equidistant bunch train with increasing bunch charges was shown to produce an increased transformer ratio in a structure with suitable fundamental frequency.  The production of these bunch trains can be achieved relatively easily by utilizing an array of beam splitters and delay stages.  However in contrast to other techniques relying on a single bunch, this approach requires a longer macro-bunch length which is more susceptible to increased energy spreads due to RF curvature.

The introduction of ideal transformer ratio current profiles ~\cite{bane-1985-a} provided a method to reduce the footprint of the drive bunch while maintaining a larger transformer ratio. Producing these current profiles directly from the photocathode is challenging however, especially due to the heightened presence of space charge forces at low energies which limits the presence of sharp edges or discontinuities.  Alternative mathematically smooth current profiles have been proposed more recently~\cite{lemerySmooth}, and are less prone to space charge forces.

\begin{figure}[h]
\centerline{\includegraphics[width=0.9\textwidth, keepaspectratio=true]{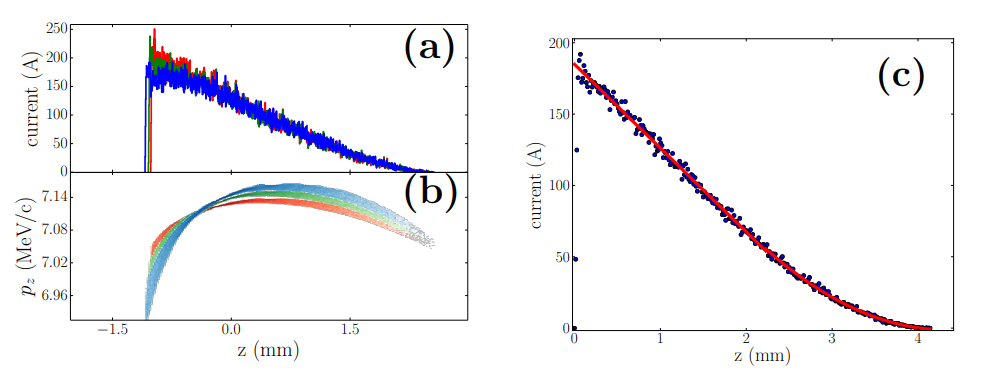}}
\caption{Evolution of the electron-bunch current (a) and longitudinal phase space (b)
along the beamline at 20 (red), 60 (green), and 100 cm (blue) from the photocathode surface and (c) comparison of the current profile numerically simulated at s = 50 cm (red trace) with a fit to equation Eq. 3.15 (blue points)\cite{lemeryThesis}. The head of the bunch is at large values of z. Figure is from Ref.~\cite{lemerySmooth}}\label{fig:laserRamp}
\end{figure}

Nonetheless, space charge forces also generally limit the charge density and preservation of current profiles emitted from the photocathode.  In Fig.~\ref{fig:laserRamp}, the evolution of a quadratic profile is shown along a distance of 1 m from the photocathode. The evolution succumbs to space charge forces which lead to increased energy spreads which elongate the bunch.  In addition, the charge density along the bunch leads to a linearization along the bunch tail.  As shown in Fig.~\ref{fig:laserRamp}(c), comparison of the bunch at z=50~cm is in good agreement with the ideal current profile; see ~\cite{lemerySmooth, lemeryThesis} for details.  Further studies and optimizations would be required to develop a usable bunch after e.g. emittance compensation and energy scaling for use in a collider.

Recent work on longitudinal shaping approaches utilize more complex methods including spatial light modulators, dispersive optics, and feedback mechanisms ~\cite{goodLaser}.  The sharp edges of the desired pulses are also practically limited by bandwidth and diffraction, see ~\cite{lemerySmooth, lemeryThesis} for further discussion. Recent demonstrations have also shown the usefulness of this scheme in an L-band photoinjector, leading to an observed transformer ratio of 4.6 in a plasma channel~\cite{pitzR}.  However the limited accelerating gradients in L-band and lower frequency RF-guns will likely require further bunch compression to realize large accelerating gradients (100+ MV/m); see~\cite{lemerySmooth} for a discussion on this topic.  In contrast, the large accelerating gradients in S- and possibly X-band guns are promising to preserve high charge densities which will be useful to support large accelerating gradients without further bunch compression. An unfortunate drawback here however is the increased RF curvature along the bunch.

\subsection{Alternative promising techniques}

The major limitation of all existing shaping methods is collective effects such as the space-charge effect or the CSR effect. Recently two shaping methods that work with a high charge bunch were proposed and were backed with rigorous simulations. One is transverse deflecting cavity (TDC) based shaping, and the other is the optimization of the entire beamline. We will shortly discuss these methods in this section.

\begin{figure}[h]
\centerline{\includegraphics[width=0.5\textwidth, keepaspectratio=true]{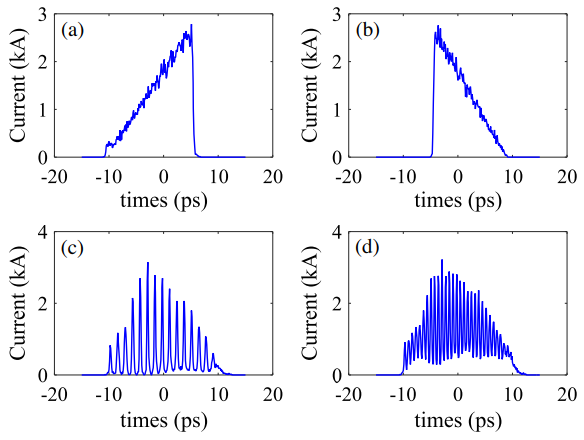}}
\caption{TDC-based longitudinal profile shaping. For a 60-nC bunch, different masks were applied to obtain (a) door-step, (b) reverse-triangle, (c) sub-ps bunch train, and (d) comb-like bunch. Charge after mask for each case was 22, 18, 12, and 23 nC, respectively. Figure is from Ref.~\cite{ha-2020-a}}\label{fig_sec32_tdcshaping}
\end{figure}

TDC is a beamline element that can introduce time and transverse coordinate correlation ($x$ or $y$). Thus, if the initial transverse components are minimized by focusing quadrupole magnets, the time component projected to the transverse plane by TDC can dominantly determine the transverse profile. If the mask is applied at this location, the mask can shape the temporal profile of the bunch. Here the remained correlation can be eliminated by the following beamline (e.g., a quadrupole magnet followed by a TDC or two to three TDCs). Ref. \cite{ha-2020-a} used this concept for arbitrary profile shaping and demonstrated its capability of high-charge shaping in simulation. They used a 60 nC bunch and generated several different current profiles as shown in the figure. Here the door-step profile, which can provide a high transformer ratio, has a total charge of 22 nC (see Fig.~\ref{fig_sec32_tdcshaping}). Because manipulation does not require any dispersive element, CSR does not affect the manipulation process. The space-charge effect is also insignificant at large beam energies.

While the method is attractive due to its CSR-free feature, the mask is its critical weak point. As mentioned in the earlier sections, charge loss by the mask may introduce various issues. However, the use of the mask is necessary for TDC-based shaping. Currently, one way to minimize such a charge-loss issue is pre-shaping \cite{ha-2020-a}. If the emission-based shaping provides coarse pre-shaping for high charge bunches, the transmission for TDC-based shaping can easily reach 70\% or higher. Thus, it would be necessary to continue the research on combining various shaping methods most effectively.

\begin{figure}
\centerline{\includegraphics[width=0.8\textwidth, keepaspectratio=true]{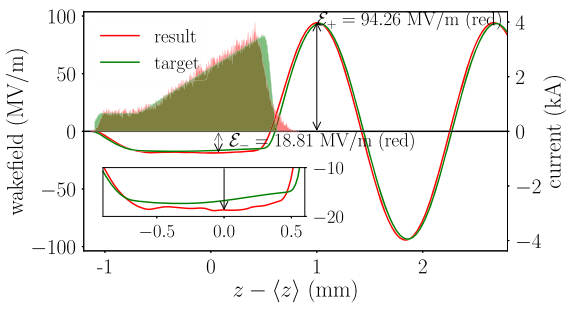}}
\caption{Longitudinal shaping by emission-based shaping with entire beamline optimization. Green and red colors represent target and simulation results, respectively. Shaded profiles are target and simulated current profiles, and curves are corresponding wakefields. Figure is from Ref.~\cite{tan-2021-a}}\label{fig_sec32_optimshaping}
\end{figure}

The other proposed method is the optimization of the entire beamline \cite{tan-2021-a}. Ref.~\cite{tan-2021-a} designed a 1-GeV beamline including injector, accelerator, and compressor, which may provide a drive bunch in practical level. Figure \ref{fig_sec32_optimshaping} shows the comparison of optimization target and simulated final current profile. The drive bunch has several requirements for collinear wakefield acceleration. The requirements are triangular shape for transformer ratio, high charge for high gradient, short bunch length for high gradient, and a large positive longitudinal chirp for beam break up suppression. Simulation result in Ref.~\cite{tan-2021-a} demonstrated a drive bunch satisfying all those requirements for the first time. Here, two key technologies were (i) emission-based shaping with a long laser pulse to mitigate space-charge effect and (ii) backward-tracking based optimization with simplified physics models to find initial requirements. When the shaped laser generates a shaped longitudinal profile, the following beamline compressed and preserved the profile. Various nonlinear effects from the beamline, including CSR, compensated each other and helped preserve the shape.

Although the method successfully demonstrated the generation of high-quality drive bunch, such optimization in the real machine could be extremely challenging. Also, the method does not work with damping rings, so it is not applicable to the main bunch.

\section{Structure wakefield acceleration - Comparing CWA with TBA}
Structure wakefield acceleration (SWFA) has a great potential to deliver high-quality witness beams (also referred to as main bunches). Compared with other technologies in the optical frequency range, the witness bunch accelerated by the SWFA approach occupies a small fraction of the RF cycle, so a high-luminosity main beam with a potentially lower energy spread can be delivered. This makes SWFA an attractive candidate towards a future energy-frontier linear collider.

To achieve high-efficiency high-gradient acceleration, there are two possible approaches in SWFA. The first one is to increase the transformer ratio in CWA, as discussed in Section 3. The other approach improves the energy transfer efficiency from the drive beam to the witness beam by decoupling the RF structure designs for the drive and witness beams, in the TBA scheme. 

Between the two schemes of SWFA, TBA has the following advantages as an AAC candidate towards a TeV collider. In general, TBA provides more degrees of freedom to meet the often very different requirement for the drive and witness beam.

 The TBA scheme offers more flexible beamline tuning and selection of the beam structure. As stated in Section 2.1, in TBA, the drive beam often has a high charge (ranging from a few nC to a few hundred nC, depending on the frequency band and type of application). The witness beam, or the main beam, is often a low-emittance beam which carries lower charge. Their beam energies are often drastically different. The main beam will be eventually accelerated to a few hundred GeV or a few TeV, while the drive beam gets decelerated from a lower initial energy. In the CLIC CDR, the drive beam initial energy is 2.4 GeV. In the current AWA setup, the drive beam starts from 70 MeV. The difference in charge and energy makes beamline optics optimization a great challenge if both beams travel in the same structures, as in the CWA scheme.

\begin{figure}[htb]
   \centering
   \includegraphics*[width=.9\columnwidth]{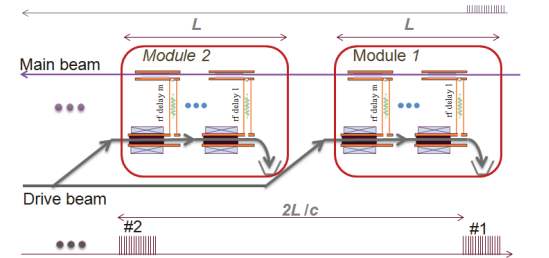}
   \caption{Schematic of TBA in two stages. (Reused from Ref.~\cite{JingIPAC13}).\label{fig:staging}}
\end{figure}

One critical technology to transition a structure prototype into a modular design for large-scale applications, like linear colliers, is the staging acceleration with sequential modules of power extractors on the drive beamline and accelerators on the main beamline. Figure~\ref{fig:staging} shows the schematic of staging in TBA. Two drive beam trains are separated with a fast kicker, and set to two pairs of power extractors. The generated wakefield is then transferred to Module 1 and Module 2 accelerators, where the main beam gets accelerated sequentially. Staging demonstrations have been completed at AWA~\cite{jingNIMA} for example. The beam synchronization condition poses less challenge for TBA compared to CWA.

TBA also provides flexibility in RF structure designs. In contrast to the CWA scheme, where deceleration of the drive beam and acceleration of the witness beam happen in the same structure, TBA scheme uses two parallel structures. The RF structure used on the drive beamline, as the power extractor structure, and the structure used on witness beamline, as the accelerator structure, can be optimized separately in the TBA scheme. To improve the energy transfer efficiency, power extractor structures are often designed to have a high group velocity, which helps with extracting high-power wakefield pulses from the drive beam. The accelerator structures often have a higher shunt impedance in comparison, to provide a high accelerating gradient for the main beam. In iris-loaded accelerating cavities, there is a trade-off between the group velocity and the shunt impedance; the above requirements mean that power extractor structures have a larger beam aperture than the accelerator structures. Advanced structure designs have been proposed, such as metamaterial structures~\cite{mtm}, dielectric disk-loaded structures~\cite{dda}, and photonic bandgap structures~\cite{pbg}, which could better serve the needs of SWFA structures requirements.

Bunch shaping techniques could greatly help both CWA and TBA to achieve high-gradient high-efficiency acceleration. Section 3 discussed how longitudinal bunch shaping could improve CWA performance. For TBA, proper longitudinal shaping could generate a more uniform in the drive bunch, to increase the interaction length, and could also reduce the energy spread in the main beam, to improve its luminosity.

\section{Conclusion}
Beam driven acceleration in structures is a relatively simple way to transfer energy between two electron beams with attractive accelerating gradients.  The simplicity of dielectric, corrugated, bimetalic, or other exotic structures is a large advantage over plasma-based sources which require differential pumping, lasers or high-voltage discharges for operation.  

We have discussed available high power electron sources which could provide the forefront of a CWA collider.  The available and extensively tested L and S-band technologies can provide means toward two different regimes, especially the very high charge regime using up to 100~nC with L-band; this approach could be used to drive wakefields with lower frequencies to mitigate instabilities. Alternatively, the large gradients supported by S-band allow the production of low-emittance beams which could be used to drive wakes in structures with smaller apertures, potentially providing larger accelerating gradients.  Both technologies can support bunch trains, and extensive simulations should be conducted to compare both approaches in supporting the development of a collider.

We also discussed various shaping approaches which have been proposed and demonstrated.  The previous two decades has led to significant progress in this field, enabling the production of current profiles which can drive ideal wakes in suitable structures.  Further optimizations and demonstrations will be required to select the appropriate shaping technique to enable the development of a TeV+ collider, especially with considerations on the large required beam powers.

While our discussion has primarily focused on CWA, we have also pointed out that the appealing differences that TBA brings forward.  While CLIC has pushed this technology with cavity based accelerators and decelerators, there is also progress from ANL on slow-wave dielectric or corrugated-based decelerators.  More work is needed on CWA to compete with existing TBA designs especially with regards to beam generation and shaping, transport, staging, and timing.


\bibliographystyle{unsrt}
\bibliography{References}

\end{document}